\documentclass[10pt,letterpaper]{article}
\usepackage{cogsci}

\cogscifinalcopy % Uncomment this line for the final submission 
\usepackage[utf8]{inputenc}
\usepackage[T1]{fontenc}
\usepackage[
    backend=biber,
    style=apa,
    natbib=true,
    doi=false,
    isbn=false,
    url=false,
]{biblatex}
\usepackage{xcolor}
\usepackage{amsmath}
\usepackage{graphicx}
\usepackage{multirow}
\usepackage{makecell}
\usepackage{tabularx}
\usepackage{adjustbox}
\usepackage{pslatex}
\usepackage{float}
\usepackage[none]{hyphenat} % Sometimes it can be useful to turn off
\title{Integrating Radiomics with Deep Learning Enhances Multiple Sclerosis Lesion Delineation}
 
\author{
\textbf{Nadezhda Alsahanova$\dagger$, Pavel Bartenev$\dagger$, Maksim Sharaev (m.sharaev@skoltech.ru)} \\
  Skolkovo Institute of Science and Technology, Moscow, 121205, Russia
 \\
BIMAI-Lab, Biomedically Informed Artificial Intelligence Laboratory, University of Sharjah, Sharjah, United Arab Emirates \\
  \AND \textbf{Milos Ljubisavljevic, Taleb Al. Mansoori, and Yauhen Statsenko  (e.a.statsenko@uaeu.ac.ae)} \\
  College of Medicine and Health Sciences, United Arab Emirates University \\
  Al Ain city, Abu Dhabi Emirate, 15551 United Arab Emirates
  \\
  \bf $\dagger$ These authors contributed equally to this work.
}
\begin{document}

\maketitle
% \newpage

    \begin{abstract}
    % Multiple sclerosis (MS) diagnosis requires precise lesion segmentation, where current deep learning approaches face robustness challenges. 
    % This study improves MS lesion segmentation in FLAIR MRI by fusing radiomic features with imaging data in an attention-augmented U-Net architecture. 
    % We analyzed scans of 46 patients (1102 slices), extracting radiomics features after intensity normalization.
    % Our radiomics-enhanced model demonstrated superior performance to conventional methods, achieving a Dice score of $0.78 \pm 0.05$ ($p<0.001$) in delineating white matter lesions. 
    % The integration of quantitative radiomics features with raw image data significantly boosted segmentation accuracy while maintaining computational efficiency. 
    % These results address critical limitations of current MAGNIMS criteria by enabling more reliable demonstration of lesion dissemination in time and space.
    % The proposed approach shows potential to support clinical decision-making in MS diagnosis and monitoring. 
    % Future validation should assess generalizability across larger, multi-center cohorts. 
    % This work advances AI-assisted neuroimaging by successfully combining radiomics and deep learning for improved MS lesion characterization
    \textbf{Background}:
    Accurate lesion segmentation is critical for multiple sclerosis (MS) diagnosis, yet current deep learning approaches face robustness challenges.
    \\
    \textbf{Aim}:
    This study improves MS lesion segmentation by combining data fusion and deep learning techniques. 
    \textbf{Materials and Methods}:
    We suggested novel radiomic features (concentration rate and Rényi entropy) to characterize different MS lesion types and fused these with raw imaging data.
    The study integrated radiomic features with imaging data through a ResNeXt-UNet architecture and attention-augmented U-Net architecture.
    Our approach was evaluated on scans from 46 patients (1102 slices), comparing performance before and after data fusion.
    \\
    \textbf{Results}:
    The radiomics-enhanced ResNeXt-UNet demonstrated high segmentation accuracy, achieving significant improvements in precision and sensitivity over the MRI-only baseline and a Dice score of 0.774$\pm$0.05; $p<0.001$ according to Bonferroni-adjusted Wilcoxon signed-rank tests. 
    The radiomics-enhanced attention-augmented U-Net model showed a greater model stability evidenced by reduced performance variability (SDD = 0.18 ± 0.09 vs. 0.21 ± 0.06; $p=0.03$) and smoother validation curves with radiomics integration.
    \\
    \textbf{Conclusion}:
    These results validate our hypothesis that fusing radiomics with raw imaging data boosts segmentation performance and stability in state-of-the-art models.
    % The fusion of quantitative radiomics with raw image data enhanced segmentation accuracy while maintaining computational efficiency, addressing key limitations of MAGNIMS criteria by enabling more reliable assessment of lesion dissemination in time and space.
    % These results highlight the clinical potential of combining radiomics and deep learning for MS diagnosis and monitoring. Future studies should validate generalizability across larger, multi-center cohorts. This work advances AI-assisted neuroimaging by providing a robust framework for improved MS lesion characterization, with implications for personalized treatment planning.
    % These results validate our hypothesis that fusing radiomics with raw imaging data boosts segmentation performance and stability in state-of-the-art models. 
    %
    %The proposed framework addresses key limitations of current MAGNIMS criteria by enabling more reliable assessment of lesion dissemination. 
    %This work advances AI-assisted neuroimaging by demonstrating the clinical potential of combining radiomics and deep learning for MS diagnosis and monitoring, with implications for personalized treatment planning. 
    %Future studies should validate generalizability across multi-center cohorts.
    
                        
    \end{abstract}
% \newpage
% \onecolumn
\section*{Acronyms}

    {The following abbreviations are used in this manuscript:\\
        \noindent 
        \begin{tabular}{@{}ll}
        AM      &   attention mechanism                              \\
        CNN     &   convolutional neural network                     \\
        CR      &   concentration rate                               \\
        FLAIR   &   fluid-attenuated inversion recovery              \\
        MAGNIMS &   magnetic resonance imaging in multiple           \\
                &   sclerosis                                        \\
        ML      &   machine learning                                 \\
        MS      &   multiple sclerosis                               \\
        qMRI    &   quantitative MRI                                 \\  
        RE      &   Rényi entropy                                    \\ 
        SDD     &   standard deviation of derivatives                \\   
        SOTA    &   state-of-the-art
        \end{tabular}
    }

\section{Introduction}
    
    Multiple Sclerosis (MS) is a chronic autoimmune disorder that affects approximately 2.8 million individuals worldwide, making it one of the most prevalent neurological diseases among young adults (\cite{sadeghibakhi2022ms}). 
    It presents with nonspecific symptoms, e.g.,  visual disturbances due to optic neuritis, numbness or tingling as sensory symptoms, fatigue, motor weakness and coordination problems. 
    Therefore, early diagnostics of the disease requires advanced image techniques (brain MRI with contrast enhancement) and laboratory tests of cerebrospinal fluid (\cite{statsenko2023multimodal}). 
    % \textcolor{blue}{A major challenge in MS lesion visual detection stems from the high heterogeneity of lesions, which vary significantly in contrast, spatial distribution, and size.}
    % %(Is it correct?)  
    % \textcolor{green}{NO the heterogeneity of disease phenotypes (CLINICAL forms) makes research on MS chalenging. As for the "heterogeneity of lesions", I would say that this assumption looks wired: it is hard to understand what does it mean - heterogenious shape? Size? location? - none of this is specific for MS}
    AI-enhanced neuroimaging has emerged as a promising tool to support clinical decision-making, offering potential improvements in early detection and monitoring (\cite{siddiqui2024role, onciul2025artificial}).  

    Machine learning (ML), particularly convolutional neural networks (CNNs), has demonstrated significant success in automating the detection of brain pathologies (\cite{aliev2021convolutional,rondinella2023boosting}). 
    However, challenges persist in model robustness and the effective integration of lesion-specific knowledge into deep learning frameworks. 
    Radiomics holds promises to address these challenges and enhance ML-based diagnostics by extracting high-dimensional quantitative features from medical images.
    These features quantify structural heterogeneity and  cover such characteristics as texture, shape, and intensity patterns.
    Recently, radiomics was successfully applied to detect mild cognitive impairment (MCI) by identifying volumetric changes in brain structures (\cite{zubrikhina2023machine}). 
    Similarly, our prior work demonstrated that combining radiomics with neural networks improves focal cortical dysplasia detection by capturing subtle pathological features often missed in visual assessments (\cite{alsahanova2025knowledge}).  

    Attention-augmented U-Nets, have achieved notable performance in MS lesion segmentation, therefore they are considered state-of-the-art (SOTA) approaches  (\cite{rondinella2023boosting}). 
    The authors validated SOTA performance through extensive experiments, comparing it against other leading methods on benchmark datasets such as the MICCAI MS lesion segmentation challenge data. 
    Their results demonstrated superior segmentation accuracy (e.g., higher Dice scores), confirming  SOTA status of their model. 
    %Despite these advances, further improvements in accuracy and robustness are needed, particularly for heterogeneous lesion patterns.  ------ IT IS HARD TO SAY, WHAT "heterogenious lesion pattern means" 
    Still, the search for more accurate and robust models remains actual.

    The demand for accurate automatic delineation of MS lesions stems from radiologists' need for a tool to assess neuroinflammatory brain changes efficiently.
    Once developed, this tool will have broad practical implications, supporting the application of MS diagnostic criteria by objectively demonstrating the dissemination of white matter lesions in time and space. 
    High-accuracy quantitative MRI assessment is critical, as the Magnetic Resonance Imaging in Multiple Sclerosis (MAGNIMS) criteria are non-specific, underscoring the need for follow-up examinations with longitudinal comparison of imaging findings.

\section{Aims and Objectives}

    This study aims to improve the accuracy and robustness of MS lesion segmentation in MRI scans by combining data fusion and deep learning techniques. 
    \textit{Hypothetically}, fusion of imaging data with radiomical features retrieved from the same images boosts the performance of the models trained to segment multiple sclerosis lesions from FLAIR images. 
    
    To reach the study objective we formulated the following tasks:
    \begin{itemize}
        \item 
        Develop radiomical features for data fusion and compare radiomics values for different types of MS lesions  
        % \textcolor{red}{compare radiomics values for different types of MS lesions}
        % \textcolor{green}{check if you can manage to do this}
            % Develop mathematical features for the detection of signal hyperintensities on FLAIR MRI scans, the primary marker of MS lesions.
        \item 
        Assess the efficiency of ML-based segmentation before and after the data fusion
            % Train a CNN model using both extracted features and raw FLAIR images, demonstrating improved segmentation accuracy compared to models trained on FLAIR alone.    
        \item 
        Test whether fusing radiomics with raw images improves the stability and performance of a recent SOTA model for lesion segmentation. 
            % Implement a state-of-the-art (SOTA) model, proving that combining FLAIR scans with additional features improves model stability and segmentation performance.
    \end{itemize}

\section{Materials and Methods}

    \subsection{Study Cohort and Data Preprocessing}
        
\begin{figure}[h]
          \centering
          \includegraphics[width=0.49\textwidth]{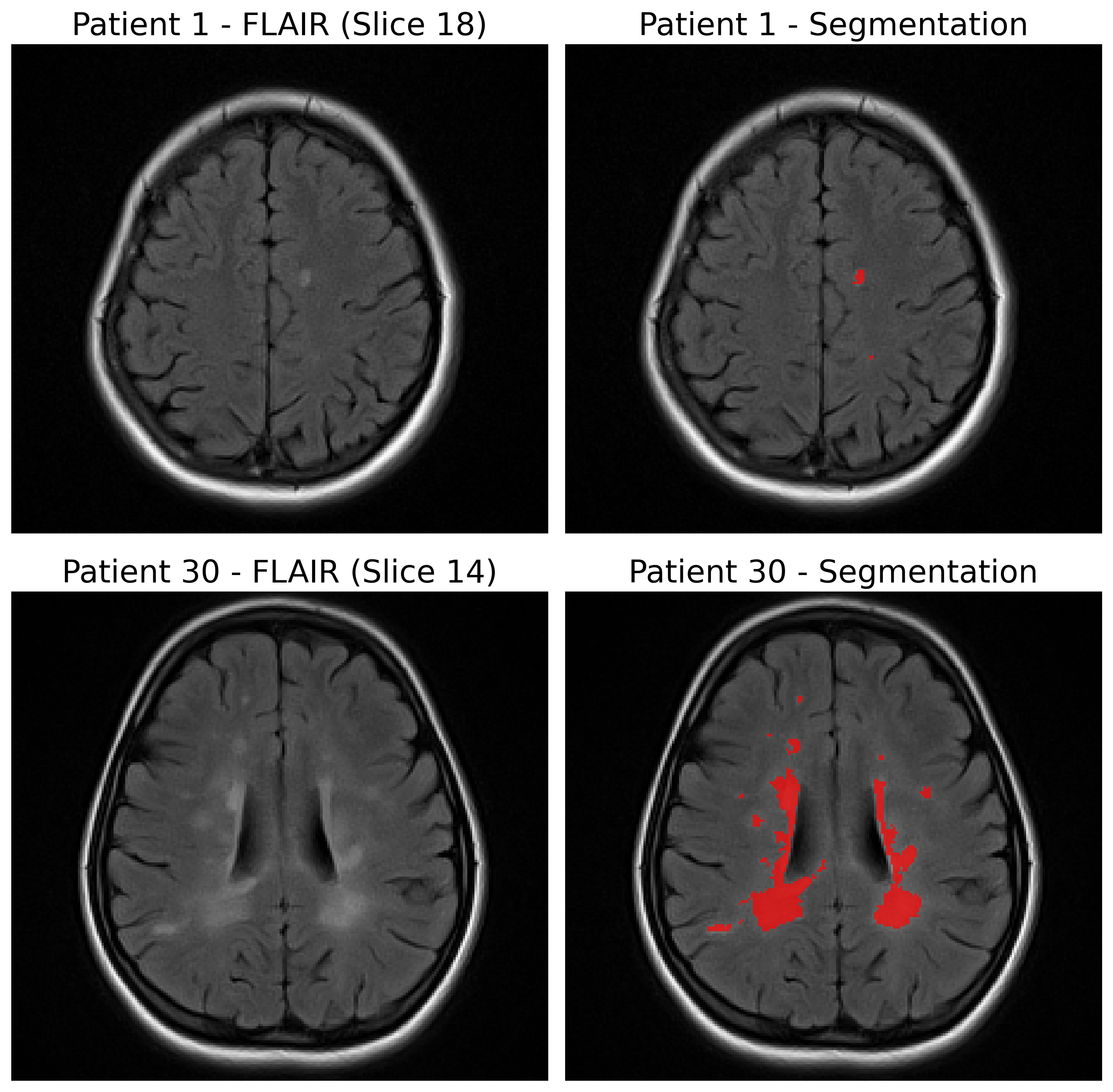}
          \caption{Examples of FLAIR slices and segmentations.}
          \label{fig:1}
        \end{figure}

        The study dataset included results MRI findings of the brain of 46 patients treated for MS in Tawam Hospital, Al Ain city. 
        From the PACS server, our team collected diagnostic images of FLAIR sequence -- the primary diagnostic modality for detection and delineation of white matter lesions in MS.
        The images had isometric voxel with the size ranging from 0.8 to 1~mm. 
        Figure~\ref{fig:1} shows FLAIR slices with white matter lesion segmentations.
        The models were trained and tested on the labeled MS lesion dataset containing 1102 slices.
        Prior to the extraction of radiomical features, we performed intensity normalisation. 

    \subsection{Research Methodology}

        \textit{To complete the first task}, we resorted to the same radiomical features as in our recent study on focal cortical dysplasia. 
        \textbf{Concentraion rate} (CR) was the most promising feature to use because which captures local hyperintensities while being robust to extreme outliers by excluding the highest values.
        Equation~\ref{eq:1_CR} describes a way to compute CR for a pixel at position $(i,j)$ with scanning window $W_s(i,j)$ of size $2s+1 \times 2s+1$ where $s=2$.   
        In the equation, 
        $num$ is a number of high-intensity pixels to sum, 
        $m$ is a number of highest pixels to exclude avoiding outliers, and
        $X_{(k)}$ is the $i$-th highest order statistic of gray scales of $X$ in the scanning window $W_s(i,j)$.
        Herein, $N$ is the total number of pixels in the neighborhood scanning window computed according to Equation~\ref{eq:2_N}. 
        
        \textbf{Rényi entropy} (RE) was a top performing feature for discovering hyperintense lesions in the white and gray matters in our previous research,  which provided rationale to apply RE to the current study (\cite{alsahanova2025knowledge}). 
        RE of order $\alpha>0, \; \alpha\neq 1$  is a measure of ``disorder'' of gray scales (\cite{Renyi60,Renyi61}). 
        To calculate it, we used the standard gray-level co-occurrence matrix (GLCM) which records how often pairs of voxels with specific intensity appear at a given distance and direction in the image (\cite{Sur12,Chit19}).
        RE is calculated for each element $f_{k,l}$  in $256^2$ GLCM matrix as per Equation~\ref{eq:3_RA}.
        We computed GLCMs for distances 1 and 2 pixels across 4 primary directions within the scanning window $W_s(i,j)$ (\cite{GLCM}). 
        Then, the summation of GLCM matrices produced a combined matrix. 
        
        \begin{equation}
            \label{eq:1_CR}
            \text{CR}(i,j) = \sum_{k=N-\text{num}-m+1}^{N-m} X_{(k)}
        \end{equation}

        \begin{equation}
            \label{eq:2_N}
            N = (2s+1)^2
        \end{equation}
        
        \begin{equation}
            \label{eq:3_RA}
            H_\alpha(i,j)=(1-\alpha)^{-1} \log \left(\sum\limits_{i,j=0}^{255}f_{k,l}^\alpha \right)
        \end{equation}

        To compute CR, we used the parameters $s = 2$, $num = 15$, and $m = 5$ in Equations~\ref{eq:1_CR} and ~\ref{eq:2_N}. 
        For Equation~\ref{eq:3_RA}, we set $s = 5$ and $\alpha = 7$ to calculate RE.

        \textit{Working on the second task}, our research team trained a CNN model to segment MS lesions in FLAIR images of the brain.
        At the input to the model, we used either raw FLAIR images or their combination with the extracted radiomical features. 
        The performance metrics were dice score, precision and sensitivity. 
        The architecture of the CNN model is in the next subsection. 
        
        \textit{To accomplish the third task}, we implemented an attention-augmented U-Net model in our study. 
        According to \cite{rondinella2023boosting}, the model reached reputable performance in MS lesion segmentation task.
        The author validated SOTA performance through extensive experiments, comparing it against other leading methods on benchmark datasets such as the MICCAI MS lesion segmentation challenge data. 
        Their results demonstrated superior segmentation accuracy (e.g., higher Dice scores), confirming its SOTA status. To make it suitable for our research, we slightly modified the architecture by removing the LSTM block from the bottleneck of the U-Net, since our dataset consists of non-adjacent MRI slices, making the block redundant.  Then the model was trained on raw FLAIR images and a combination of them with radiomical features. Cross-validation was performed using the same metrics as for previous task to access the results.

        To assess the stability of model training, we calculated the standard deviation of derivatives (SDD) which reflects the smoothness of a validation curve (see Equation~\ref{eq:4_SDD}).
        In this formula, $N$ is the number of validations performed during training and $d_i$ represents the change in the metric between consecutive validations (see Equation~\ref{eq:5_d}). 

        \begin{equation}
            \label{eq:4_SDD}
            SDD = \sqrt{\frac{1}{N} \sum_{i=1}^N(d_i - \overline{d})^2}
        \end{equation}

        \begin{equation}
            \label{eq:5_d}
            d_i = val\_score_i - val\_score_{i-1}
            % SDD = \sqrt{\frac{1}{N} \sum_{i=1}^N(d_i - \overline{d})^2}
        \end{equation}

    \subsection{Model Architecture} 
    
        \textbf{The CNN model} in the second part of our study followed a ResNeXt-UNet architecture for semantic segmentation.
        The architecture combines the robust feature extraction capabilities of ResNeXt-50 (\cite{xie2017aggregated}) with the segmentation ability of the U-Net encoder-decoder framework. 
        Recently, this combination demonstrated high segmentation quality in detecting brain tumors (\cite{rai2021automatic}).

        The network consists of three main components: an encoder backbone, a decoder pathway, and skip connections for multi-scale feature fusion.
        The encoder utilizes a ResNeXt-50 backbone pre-trained on ImageNet, providing strong transferable feature representations. 
        ResNeXt-50 employs grouped convolutions with cardinality of 32 and base width of 4, offering improved representational capacity compared to standard ResNet architectures while maintaining computational efficiency.    
        The decoder pathway consists of four upsampling blocks (DecoderBlock) that progressively reconstruct the spatial resolution while reducing feature dimensionality.
        The final classification head consists of two convolutional layers: a 3×3 convolution reducing channels from 256 to 128, followed by a final 3×3 convolution mapping to the desired number of output classes. 
        The network outputs raw logits for each pixel, with subsequent softmax activation applied during inference.
        All convolutional layers except the final output layer are followed by ReLU activation functions.

        \textbf{The attention-augmented U-Net model} employs a Tiramisu-style (\cite{jegou2017}) Fully-Convolutional DenseNet that keeps the classic U-Net encoder–decoder layout: five down-sampling stages, a bottleneck, and five mirrored up-sampling stages linked by skip connections. Squeeze-and-Attention blocks (\cite{zhong2020}) follow every Dense Block in both encoder and decoder.

    \subsection{Training Settings}
    
        Both models were evaluated using six-fold cross-validation (8 subjects per fold for 1-4 folds and 7 subjects per fold for 5-6 folds). 

        \textbf{The CNN model} was trained using a combined loss function with equal weighting ($\gamma = 0.5$) between Dice loss and binary cross-entropy loss. Optimization was performed with AdamW, configured with an initial learning rate of $1 \cdot 10^{-4}$ and a weight decay of $1 \cdot 10^{-5}$.
        To improve convergence, a learning rate scheduling strategy was implemented. This began with a 100-iteration warmup phase, during which the learning rate linearly increased from $1 \cdot 10^{-6}$ to the base value. After warmup, cosine annealing was applied across subsequent epochs, gradually reducing the learning rate to a minimum of $5 \cdot 10^{-6}$. This approach stabilized training dynamics and enhanced final model performance.
    
        Input data were prepared by extracting 2D slices from 3D FLAIR volumes, resizing each slice to $256 \times 256$ pixels. Intensity normalization was applied per slice using min-max scaling to standardize input ranges. During training, data augmentation included random horizontal flipping with a 50\% probability.  
    
        Model for each fold was trained for 50 epochs using a batch size of 32.
          
        During inference, FLAIR images underwent identical preprocessing as during training. The model outputs were processed with a sigmoid activation function before thresholding, and predicted masks were resized to original dimensions using nearest-neighbor interpolation to maintain sharp lesion boundaries.  

        \textbf{The attention-augmented U-Net model} was trained using Dice loss with RMSprop optimizer. The initial learning rate was set to $1\cdot 10^{-4}$ and learning rate decay was implemented with decay factor of $0.995$ applied after every epoch. The data preprocessing and augmenations were done in similar way to the CNN model described above except that the slices were resized to $224 \times 224$ pixels. The model was trained for 40 epochs for each fold with batch size of 6.

\section{Results}
    \subsection{Radiomic features}     
    
\begin{figure}[h]
          \centering
          \includegraphics[width=0.49\textwidth]{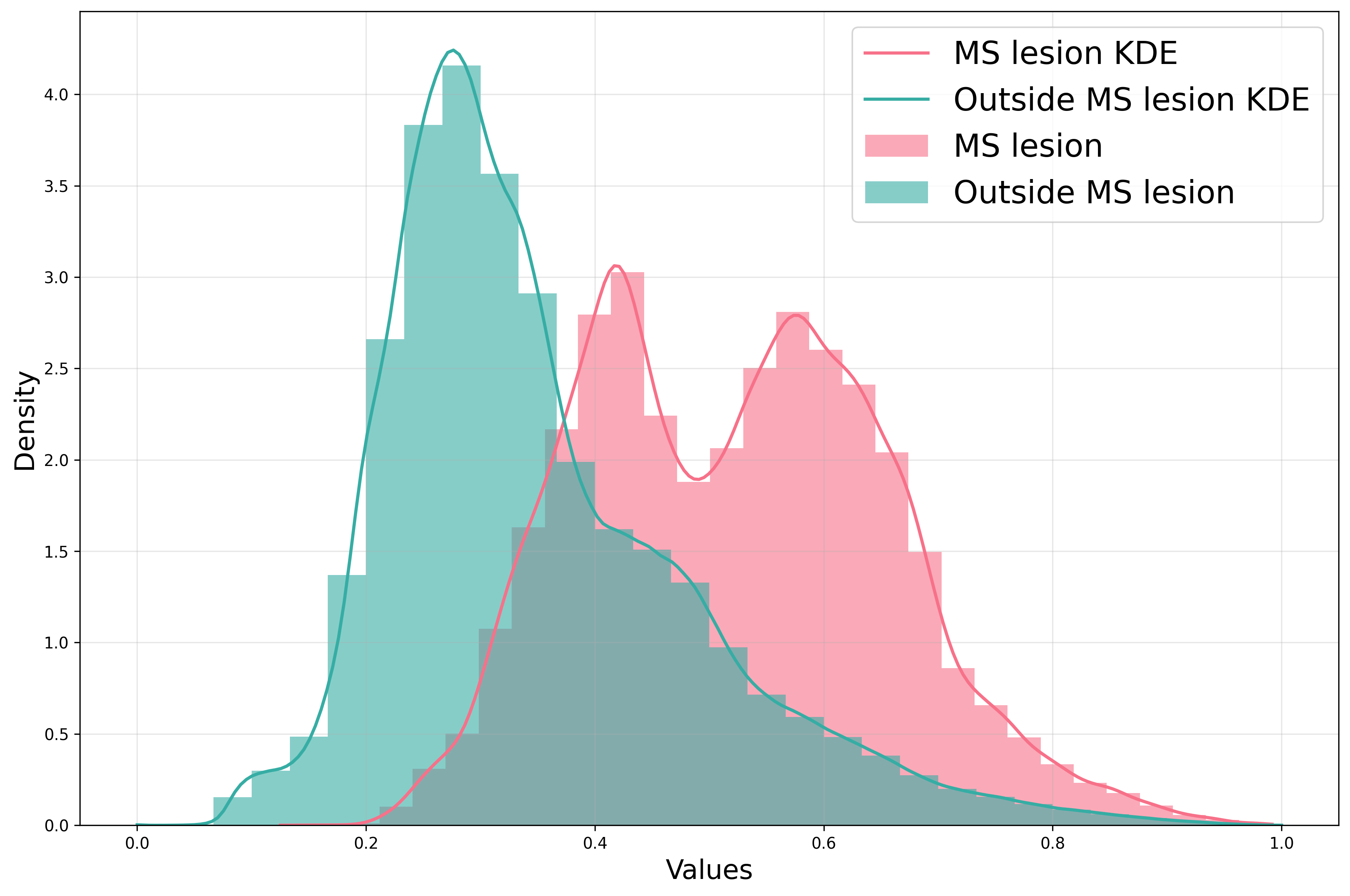}
          \caption{Values of concentration rate inside and outside white matter lesions in multiple sclerosis.
          %Comparison of CR feature value distribution for healthy and multiple sclerosis voxels.
          }
          \label{fig:CR}
        \end{figure}

    The distribution of CR values differed between voxels inside and outside MS lesions (see Figure~\ref{fig:CR}).
    CR was markedly higher within lesion boundaries, confirming its sensitivity to hyperintense regions. 
    The bimodal distribution across MS lesions may stem from two factors. 
    First, CR depends on lesion size and scanning window placement.
    It peaks, when the window is fully within a lesion.
    Contrarily, a lower CR value comes when the window includes the voxels located close to the borders but outside the lesion.    
    Second, neuroinflammatory lesions in MS vary in shape and type.
    The difference among supratentorial, infratentorial, juxtacortical or paraventricular foci may account for the dual-peak pattern of the CR distribution. 

\begin{figure}[h]
          \centering
          \includegraphics[width=0.49\textwidth]{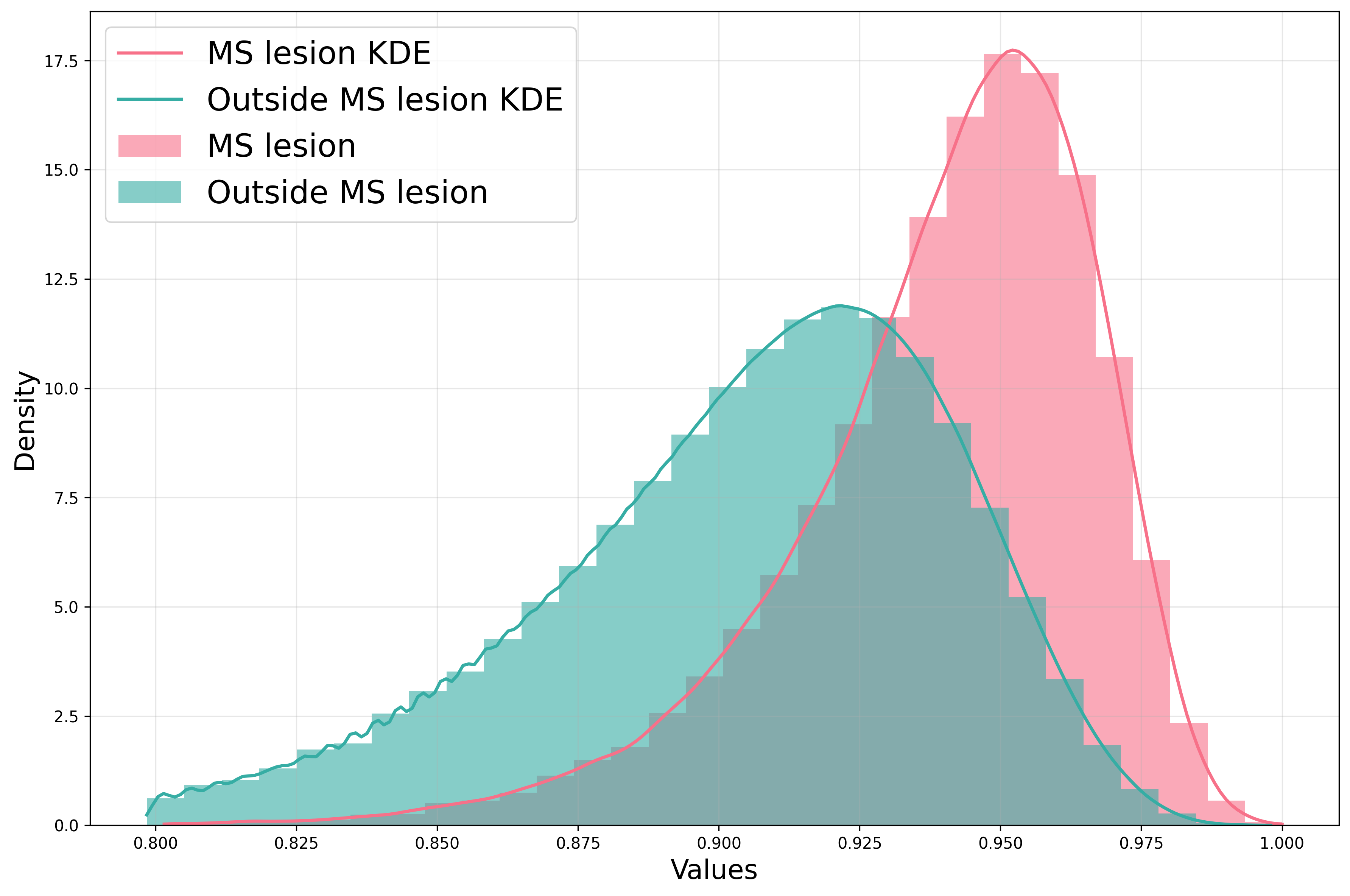}
          \caption{
          Values of  Rényi entropy inside and outside white matter lesions in multiple sclerosis.
          % Comparison of Entropy feature value distribution for healthy and multiple sclerosis voxels.
          }
          \label{fig:Entropy}
        \end{figure}

    On average, RE was also larger inside than onside MS lesions (see Figure~\ref{fig:Entropy}).
    The difference in the RE distribution evidences its capacity to detect hyperintense regions in FLAIR.
    The RE distribution across MS lesions followed a single-spike pattern.
    Hence, the patterns characterizing CR and RE in MS differ, and future studies should explore the reasons underlying this difference.
    Since CR and RE may help to identify neuroinflammation in FLAIR, we decided to use both features in data fusion described in the next subsection.
    
    % Figure~\ref{fig:Entropy} demonstrates elevated Rényi Entropy values within lesions, further validating its capacity to detect hyperintense regions. 
    % Both features effectively differentiate pathological tissue.
    
    \subsection{CNN model}
    
    Incorporating radiomics features into ML models alongside MRI data consistently improved segmentation performance across all evaluation metrics.
    The models trained on a combination of MRI and radiomical features achieved higher Dice scores, precision and sensitivity compared to those trained exclusively on MRI data (see  Table~\ref{table2}). 
    The Wilcoxon signed-rank test with Bonferroni correction for multiple comparisons showed a significant improvement in the aforementioned metrics (\cite{wilcoxon1992individual}). 
    
    The fold-by-fold analysis further confirmed the validity of the study hypothesis (see Table~\ref{table1}). 
    With the data fusion technique, we managed to elevate Dice metric across all six cross-validation folds.
    The improvement was most remarkable in folds 3 (0.77 vs 0.74), 5 (0.73 vs 0.69), and 6 (0.73 vs 0.69). 
    The consistent gain in performance suggest that the radiomics provides complementary information enhancing the neural network's segmentation.

    % Across all six cross-validation folds, the MRI+features model outperformed the MRI-only baseline in Dice score, with particularly notable improvements in folds 3 (0.77 vs 0.74), 5 (0.73 vs 0.69), and 6 (0.73 vs 0.69). 
    % The consistent performance gains across different data splits suggest that the radiomics features provide complementary information that enhances the neural network’s segmentation.

    \begin{table}[h]
    \caption{
    Metrics averaged across all cross-validation folds
    % Mean performance metrics across all cross-validation folds, with Bonferoni corrected p-values from the Wilcoxon signed-rank test comparing the CNN trained on MRI + radiomics features versus the CNN trained on MRI alone.
    }
    \label{table2}
    \begin{center}
        \adjustbox{width=0.5\textwidth}{
        \begin{tabular}{lccc}
        \hline
        \multicolumn{1}{c}{\multirow{2}{*}{\textbf{Metrics}}} 
        & \multicolumn{1}{c}{\multirow{2}{*}{\textbf{MRI}}} 
        & \multicolumn{1}{c}{\multirow{2}{*}{\makecell{\textbf{MRI +}\\\textbf{radiomics features}}}} 
        & \multicolumn{1}{c}{\multirow{2}{*}{\textbf{p-value}}}
        \\[2pt]
        \\
        \hline%\rule{0pt}{12pt}
        Dice score  & 0.680 $\pm$ 0.041 
                    & 0.706 $\pm$ 0.046
                    & $ < 0.001$  \\
        Precision   & 0.765 $\pm$ 0.045
                    & 0.774 $\pm$ 0.050
                    & $< 0.001$ \\
        Sensitivity & 0.686 $\pm$ 0.038
                    & 0.719  $\pm$ 0.041
                    &  $< 0.001$ \\[2pt] 
        \hline
         \multicolumn{4}{l}{
         \makecell[l]{
         Bonferoni-adjusted p-value from the Wilcoxon signed-rank \\
         test indicates measured difference between metrics of CNN \\
         trained on MRI with and without radiomics features
        }
         }
        \end{tabular}
        }
        \end{center}
    \end{table}

    \begin{table}[h]
    \caption{Mean Dice score for test set in each fold}
    \label{table1}
    \begin{center}
    \begin{tabular}{rllllll}
        \hline
        Experiment & 1 & 2 & 3 & 4 & 5 & 6 \\[2pt]
        \hline\rule{0pt}{12pt}
        Only MRI  & 0.67 & 0.60 & 0.74 & 0.68 & 0.69 & 0.69 \\
        MRI+features  & 0.68 & 0.62  & 0.77 & 0.71  & 0.73 & 0.73  \\[2pt]
    \hline
    \end{tabular}
    \end{center}
    \end{table}

    \subsection{U-Net with attention model}

    The application of radiomics features also boosted the segmentation accuracy of the U-Net with attention model.
    The mean Dice, Precision, and Sensitivity scores rose consistently  across all folds after enhancing the input with radiomics features (see Table~\ref{table4}).
    The result of the Wilcoxon signed-rank test adjusted for multiple comparisons evidenced a pronounced enlargement of the referenced metrics.
    In most cases, the model trained on a combination of MRI and radiomics features outperformed the baseline model, which clearly suggests the benefits of incorporating radiomics (see Table~\ref{table3}).

    %% I SHIFTED A PARAGRAPH TO THE METHODOLOGY SECTION

\begin{figure}[h]
      \centering
      \includegraphics[width=0.49\textwidth]{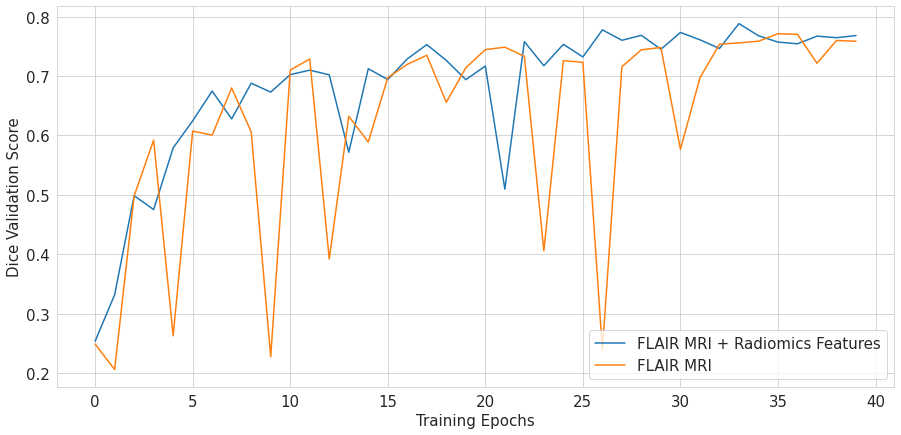}
      \caption{Dice validation scores obtained after each training epoch. 
      }
      \label{fig:val_plot}
    \end{figure}

    From the analysis of validation curves, model training was more stable after we combined data fusion and deep learning techniques:   $SDD=0.21\pm0.06$ vs $0.18\pm0.09$. 
    The result of the Wilcoxon signed-rank test indicated a noticeable drop in SDD with $p=0.03$, proving that the model enhancement with radiomics was beneficial in terms of training stability. 
    The same was clear from the visual appearance of validation curves: they were smoother with radiomics features and sharper without them  (see Figure~\ref{fig:val_plot}).

    \begin{table}[h]
    \caption{
    Metrics averaged across all cross-validation folds
    % Mean performance metrics across all cross-validation folds, with Bonferoni corrected p-values from the Wilcoxon signed-rank test comparing the U-Net with attention trained on MRI + radiomics features versus the model trained on MRI alone.
    }
    \label{table4}
    \begin{center}
        \adjustbox{width=0.5\textwidth}{
        \begin{tabular}{lccc}
        \hline
        \multicolumn{1}{c}{\multirow{2}{*}{\textbf{Metrics}}} 
        & \multicolumn{1}{c}{\multirow{2}{*}{\textbf{MRI}}} 
        & \multicolumn{1}{c}{\multirow{2}{*}{\makecell{\textbf{MRI +}\\\textbf{radiomics features}}}} 
        & \multicolumn{1}{c}{\multirow{2}{*}{\textbf{p-value}}}
        \\[2pt]
        \\
        \hline%\rule{0pt}{12pt}
        Dice score  & 0.707 $\pm$ 0.054
                    & 0.719 $\pm$ 0.063
                    & $ < 0.001$  \\
        Precision   & 0.685 $\pm$ 0.104
                    & 0.753 $\pm$ 0.058
                    & $< 0.001$ \\
        Sensitivity & 0.717 $\pm$ 0.063
                    & 0.756 $\pm$ 0.089
                    &  $< 0.001$ \\[2pt] 
        \hline
         \multicolumn{4}{l}{
         \makecell[l]{
         Bonferoni-adjusted p-value from the Wilcoxon signed-rank \\
         test indicates measured difference between metrics of CNN \\
         trained on MRI with and without radiomics features
        }}
        \end{tabular}
        }
    \end{center}
    \end{table}
    
    \begin{table}[h]
    \caption{Mean Dice score for test set in each fold}
    \label{table3}
    \begin{center}
    \begin{tabular}{rllllll}
        \hline
        Experiment & 1 & 2 & 3 & 4 & 5 & 6 \\[2pt]
        \hline\rule{0pt}{12pt}
        Only MRI  & 0.76 & 0.73 & 0.60 & 0.75 & 0.71 & 0.67 \\
        MRI+features  & 0.77 & 0.71 & 0.72 & 0.77 & 0.76 & 0.59  \\[2pt]
    \hline
    \end{tabular}
    \end{center}
    \end{table}

\section{Discussion}

    \subsection{Utility of Radiomical Findings for Data Fusion}

        \subsubsection{Data Fusion Techniques for Enhanced Diagnostic Performance}

        Data fusion methodologies enable the integration of imaging data with non-imaging clinical and molecular data, which provides a more comprehensive view of individual tumors and potentially improves the prediction of patient outcomes (\cite{wang2019}). 

        Multimodal data fusion has shown promise in medical imaging tasks like breast cancer prediction and glaucoma classification (Huang et al., 2020). 
        Early fusion, late fusion, and intermediate fusion represent common strategies for integrating multimodal data, each with its own advantages and limitations (\cite{wang2025}). 
        Early fusion concatenates features from different modalities at the input level, creating a high-dimensional feature vector that captures the collective information from all modalities; however, this approach may be sensitive to irrelevant or redundant features and may not effectively capture the complex interrelationships between modalities (Liu et al., 2022). 
        Late fusion, on the other hand, trains separate models for each modality and then combines their predictions using techniques such as weighted averaging or ensemble learning, allowing each modality to be processed independently and contributing its unique perspective to the final decision (Kline et al., 2022). 
        Intermediate fusion combines features at an intermediate stage of the processing pipeline, allowing for more flexible integration of information and potentially capturing more complex interdependencies between modalities.
        
        The increasing availability of multimodal data in healthcare, encompassing electronic health records, medical imaging, multi-omics, and environmental data, has spurred the development of AI-based data fusion techniques to enhance prediction, diagnosis, and treatment planning (\cite{mohsen2022}). 
        The fusion of multimodal brain imaging techniques is gaining popularity in mental disorder analysis, since it allows for a multidimensional analysis, providing comprehensive insights into the interrelationships between various imaging modalities (\cite{jiao2025}).
        Multimodal medical image fusion techniques coalesce multiple images from different imaging modalities to derive a fused image enriched with a wealth of information, thereby enhancing the clinical applicability of medical images (\cite{huang2020}). 
        
        The integration of data from diverse modalities, such as imaging, genomics, and clinical data, can provide a more complete picture of a patient's health, reducing the chance of misdiagnosis and improving the accuracy of diagnosis (\cite{antari2023}). 
        By integrating visual, temporal, and textual information into a unified feature representation space, a more holistic and nuanced understanding of industrial system complexities can be attained (\cite{wang2025_2}). 
        The clinical use of fusion imaging is recognized as a central component in the general scheme of clinical decision-making (\cite{zaidi2009}). 
        Overall, multimodal fusion shows significant benefits in clinical diagnosis and neuroscience research (\cite{zhang2020}).

        \subsection{Role of Radiomics in Diagnostic Imaging}

        Radiomics, an emerging field, is revolutionizing precision medicine by quantitatively analyzing medical images to extract a wealth of phenotypic features, thereby establishing a link between imaging and personalized treatment strategies (\cite{arimura2018radiomics,guiot2021review}).
        Moving beyond the traditional qualitative assessment of medical images, radiomics leverages sophisticated algorithms to mine high-dimensional data, encompassing first-, second-, and higher-order statistics, which can then be integrated with clinical and genomic information to enhance diagnostic, prognostic, and predictive accuracy (\cite{gillies2015radiomics}). 
        The radiomics pipeline typically begins with image acquisition, ensuring standardized protocols to minimize variability, followed by precise segmentation of the region of interest to delineate the anatomical or pathological structures for feature extraction (\cite{timmeren2020}). 
        Following feature extraction, the high-dimensional radiomic feature space is subjected to normalization techniques, such as Z-score standardization or min-max scaling, to mitigate the effects of differing feature scales and distributions (\cite{capobianco2020}). 
        To ensure robust and generalizable models, feature selection methodologies are crucial in radiomics, ranging from univariate statistical tests, such as t-tests and ANOVA, to advanced machine learning techniques including recursive feature elimination, least absolute shrinkage and selection operator, and tree-based methods (\cite{timmeren2020}).
        Finally, machine learning models are trained and validated to predict clinical outcomes, with careful attention to hyperparameter tuning and cross-validation strategies to prevent overfitting and assess the model's performance on unseen data (\cite{farias2021}). 
        Despite the promising applications of radiomics in diagnostic imaging, the complexities associated with data acquisition, image segmentation, feature extraction, and model validation require careful consideration to ensure the robustness and reproducibility of the results (\cite{ghuwalewala2021,kocak2019}).

        \subsection{Radiomic Features: Concentration Rate and Rényi Entropy}

        Radiomic features, including concentration rate and Rényi entropy, offer unique insights into the underlying tissue characteristics, reflecting both the distribution and complexity of voxel intensities within medical images (\cite{rizzo2018}). 
        Concentration rate, a statistical measure, quantifies the degree to which voxel intensities are clustered around a specific value, reflecting the homogeneity or heterogeneity of the tissue (\cite{yang2018}). 
        A high concentration rate suggests that a large proportion of voxels exhibit similar intensity values, indicating a more uniform tissue structure, whereas a low concentration rate implies a wider distribution of intensities, indicative of greater heterogeneity. 
        Rényi entropy, a generalization of Shannon entropy, provides a flexible measure to quantify the randomness or disorder of voxel intensities within an image, with its sensitivity adjustable via a parameter that controls the weighting of different intensity values. 
        By tuning the parameter, Rényi entropy can be tailored to emphasize either the dominant intensity patterns or the subtle variations within the image, offering a more comprehensive characterization of tissue heterogeneity.

        In the context of cancer imaging, radiomic features facilitate the non-invasive characterization of tumor heterogeneity, a critical determinant of tumor aggressiveness and therapeutic response, by quantifying the spatial distribution of voxel intensities and textural patterns within the tumor volume (\cite{mayerhoefer2020}). 
        \textbf{Entropy} measures in radiomics have been related to molecular classifications of breast cancer subtypes, with higher entropy often correlating with more aggressive subtypes (\cite{li2016}).

        Concentration rate, a measure of the degree to which voxel intensities are clustered within a specific region of interest, can provide valuable insights into the spatial distribution of metabolic activity or contrast enhancement, thereby reflecting the underlying biological processes occurring within the tissue (\cite{wang2025}). 
        Rényi entropy, a generalization of Shannon entropy, quantifies the randomness or disorder of the voxel intensity distribution, offering a more nuanced characterization of tissue heterogeneity and complexity that can be used to distinguish between different disease states and predict treatment response (\cite{tan2020}). 

        Concentration rate can be particularly useful in differentiating between benign and malignant lesions, as malignant tumors often exhibit higher concentration rates due to their increased metabolic activity and rapid cell proliferation. 
        Rényi entropy, with its ability to adjust sensitivity to different aspects of the intensity distribution through its order parameter, provides a flexible tool for capturing subtle changes in tissue texture and heterogeneity that may be indicative of early-stage disease or treatment response (\cite{cui2022}). 
        The feature enhancement capability of image fusion is visually apparent in combinations that often results in images that are superior to the original data (\cite{jiang2009}). 
        By combining concentration rate and Rényi entropy with other radiomic features and clinical data, data fusion techniques can create a more comprehensive and robust diagnostic model that is less susceptible to noise and variability in the imaging data (\cite{rasekh2024}). 
        
        The integration of concentration rate and Rényi entropy within a data fusion framework necessitates careful consideration of feature normalization, weighting, and selection techniques to ensure that each feature contributes appropriately to the final diagnostic decision. 
        Data normalization techniques, such as z-score scaling or min-max scaling, can mitigate the impact of differing feature scales and ranges, while feature weighting algorithms, such as those based on mutual information or correlation analysis, can emphasize the contributions of more informative features (\cite{hagiwara2020}). 
        The decrease of the weight entropy in a cluster illustrates the increase of certainty of a subset of features with more substantial weights in the determination of the cluster (\cite{singh2019}).
        Additionally, feature selection methods, such as principal component analysis or feature importance ranking, can identify and remove redundant or irrelevant features, further improving the model's performance and interpretability. 
        
        The application of radiomics, which involves extracting quantitative features from medical images, has shown promise in various clinical applications, including cancer diagnosis, prognosis, and treatment response prediction (\cite{liao2019}). 
        Radiomic features, such as concentration rate and Rényi entropy, can be used to characterize the heterogeneity and complexity of tumors (\cite{frank2021}). 
        Integrating the radiomic features with other modalities, like clinical and genomics data, can offer a more comprehensive view of the disease, potentially boosting the performance of diagnostic models (\cite{pfeifer2020}). 
        
        The development of fused instruments, such as PET/CT and PET/MRI stations, has bolstered the concept of complementing the strengths of one imaging modality with those of another (\cite{huang2011}). 
        While acquiring unified imaging data of good quality may be relatively straightforward in routine clinical practice, achieving this level of cooperation across different healthcare facilities represents a considerable obstacle (\cite{patyk2018}). 
        Molecular imaging significantly contributes to personalized medicine by providing noninvasive spatiotemporal information on physiological and pathological processes, which can then be used not only for accurate diagnosis and determination of the extent of disease but also for rational targeted therapy and treatment monitoring (\cite{jadvar2013}). 
        Advanced MRI techniques, quantitative methods, and artificial intelligence can evaluate brain metastases, specifically for diagnosis, including differentiating between malignancy types and evaluation of treatment response, including the differentiation between radiation necrosis and disease progression (\cite{tong2020}). By combining radiomic features like concentration rate and Rényi entropy with other data modalities through sophisticated data fusion techniques, diagnostic models can achieve enhanced performance in disease detection, characterization, and prediction (\cite{cuocolo2019,li_2020}).

\section{Conclusion}

    \begin{enumerate}
        \item %1
        The integration of concentration rate and Rényi entropy into diagnostic models allows for a more nuanced understanding of the underlying tissue characteristics, potentially improving the accuracy of diagnostic and prognostic predictions.
        These radiomic features, when combined, provide a comprehensive characterization of the imaging phenotype, capturing both the central tendency and the distributional properties of voxel intensities, which can be leveraged to improve the performance of diagnostic models. 
        \item %2
        The models trained on a combination of MRI and radiomical features achieved higher Dice scores, precision and sensitivity compared to those trained exclusively on MRI data. 
        The Wilcoxon signed-rank test with Bonferroni correction for multiple comparisons showed a significant improvement in the aforementioned metrics.
        After the incorporation of radiomics features, the enhanced model  outperformed the MRI-only baseline in Dice score, with particularly notable improvements in folds 3 (0.77 vs 0.74), 5 (0.73 vs 0.69), and 6 (0.73 vs 0.69). 
        \item %3
        Model training was more stable after we combined data fusion and deep learning techniques: $SDD=0.21\pm0.06$ vs $0.18\pm0.09$ ($p=0.03$). 
        This was apparantly seen from the validation curves that were smoother with radiomics features and sharper without them.
    \end{enumerate}

\section*{Strength and Limitations}

    The study posas several strengths.
    First, it advances the development of quantitative MRI (qMRI), which enables precise, reproducible measurements of tissue properties, moving beyond qualitative imaging.
    By providing objective biomarkers, qMRI supports precision medicine through tailored diagnoses and treatments based on individual patients’ pathophysiology. 
    For example, qMRI can stratify tumor subtypes, track neurodegeneration, or assess brain tissue health, thereby improving personalized therapeutic decisions.
    Second, the proposed approaches can be flexibly applied to various brain pathologies and other internal organs.
    Radiomics is a field of ongoing research since its emergence. 
    Bioengineers aim to interpret radiomic features and develop optimized architectures for radiomics-based models. 

    Although this study advances the understanding of radiomics, data fusion, and deep learning techniques, several limitations merit discussion.
    First, the study sample is relatively small to account for the heterogeneous clinical phenotypes of multiple sclerosis (MS). 
    However, the focus was on white matter, which exhibits a common appearance across cases.
    Second, the trained models lack sufficient reliability for clinical use, primarily because their validation was limited to a single dataset. However, data fusion techniques could potentially enhance accuracy by incorporating additional imaging sequences. Furthermore, additional research should investigate domain shift \cite{kondrateva2021domain} by applying the method to multiple independent datasets. Implementing these improvements would require both a larger patient cohort and increased computational resources.

    Given that many open multiple sclerosis datasets lack voxel-wise labels, another promising approach would be weak supervision \cite{pavlov2019weakly}, which can leverage partially labeled or image-level annotations. 
    % Once robust performance is achieved, future studies should validate the findings on an independent external dataset of MS cases

\section*{Ethics}

    The current study got approval from the local ethical committee of Tawam hospital (Al Ain) and DoH Ethical Committee (Abu Dhabi). 

\section*{Fund Acknowledgement}

    A part of the study received financial support from Joint McGill-UAEU research grant "Creating and validating generative AI models that synthesize high-quality MRI images of multiple sclerosis patients" (PI: YS, proposal 4508); NA, PB, and MS were supported by RScF grant №25-71-30008 (creating reliable Machine Learning models)

\printbibliography

\end{document}